\documentclass[reprint, nofootinbib]{revtex4-2}

\usepackage{hyperref}
\usepackage{enumerate}
\usepackage{graphicx,ulem}
\usepackage{xcolor}

\usepackage{tabularx}
\usepackage{multirow}
\usepackage{tabularray}
\usepackage{csvsimple}

\newcommand{\bwt}{\begin{widetext}}
\newcommand{\ewt}{\end{widetext}}
\newcommand{\beq}{\begin{equation}}
\newcommand{\eeq}{\end{equation}}
\newcommand{\bea}{\begin{eqnarray}}
\newcommand{\eea}{\end{eqnarray}}

\usepackage{float}
\usepackage{comment}
\usepackage{amsmath, amssymb, mathrsfs, slashed}
\usepackage{multirow, bm, IEEEtrantools}
\usepackage{acro}

\DeclareAcronym{tov}{
  short=TOV,
  long=Tolman-Openheimer-Volkoff,
}
\DeclareAcronym{sm}{
  short=SM,
  long=standard model,
}
\DeclareAcronym{ns}{
  short=NS,
  long=Neutron star,
}
\DeclareAcronym{bns}{
  short=BNS,
  long=binary neutron star,
}
\DeclareAcronym{pns}{
  short=PNS,
  long=Proto-neutron stars,
}
\DeclareAcronym{hs}{
  short=HS,
  long=hybrid star,
}
\DeclareAcronym{bh}{
  short=BH,
  long=black hole,
}
\DeclareAcronym{qcd}{
  short=QCD,
  long=quantum chromodynamics,
}
\DeclareAcronym{pqcd}{
  short=pQCD,
  long=perturbation quantum chromodynamics,
}
\DeclareAcronym{lqcd}{
  short=lQCD,
  long=lattice quantum chromodynamics,
}
\DeclareAcronym{eos}{
  short=EOS,
  long=equation of state,
}
\DeclareAcronym{nsm}{
  short=NSM,
  long=neutron star matter,
}
\DeclareAcronym{nm}{
  short=NM,
  long=nuclear matter,
}
\DeclareAcronym{ddb}{
  short=DDB,
  long=density depended couplings with Bayesian analysis,
}
\DeclareAcronym{rmf}{
  short=RMF,
  long=relativistic mean field,
}
\DeclareAcronym{ai}{
  short=AI,
  long=artificial intelligence,
}
\DeclareAcronym{gw}{
  short=GW,
  long=gravitational wave,
}
\DeclareAcronym{gr}{
  short=GR,
  long=general relativity,
}
\DeclareAcronym{nicer}{
  short=NICER,
  long=Neutron Star Interior Composition ExploreR,
}

\DeclareAcronym{ur}{
  short=UR,
  long=universal relation,
}
\DeclareAcronym{npe}{
  short={\it npe},
  long=neutron-proton-electron,
}
\DeclareAcronym{rhic}{
  short=RHIC,
  long=relativistic heavy-ion collider,
}
\DeclareAcronym{lhc}{
  short=LHC,
  long=large hadron collider,
}
\DeclareAcronym{eft}{
  short=EFT,
  long=effective field theory,
}

\begin{document}
\title{
% Evolution of an exotic proto-neutron star into black hole\\
Constraints on maximum neutron star mass from proto-neutron star evolution %\\
% Identifying the presence of hyperons from proto-neutron stars\\
}

\author{Deepak Kumar$^{1,2}$} \email{deepak.kumar@iopb.res.in}
\author{Tuhin Malik$^{3}$} \email{tuhin.malik@uc.pt}
\author{Hiranmaya Mishra$^{2,4}$} \email{hiranmaya@niser.ac.in}
\author{Constan\c ca Provid\^encia $^{3}$}

\affiliation{$^{1}$Department of Physics, Indian Institute of Science Education and Research, Bhopal, 462 066, India}
\affiliation{$^{2}$Institute of Physics, Sachivalaya Marg, Bhubaneswar 751005, India}
\affiliation{$^{3}$CFisUC, Department of Physics, University of Coimbra, PT 3004-516 Coimbra, Portugal}
\affiliation{$^4$School of Physics, National Institute of Science Education and Research, An OCC of Homi Bhabha National Institute, Jatni - 752050, India}

\date{\today}

\begin{abstract}
A proto-neutron star (PNS) gets formed after a successful supernova when the stellar remnant decouples from the ejecta. In this study, we explore a relativistic framework for the finite-temperature $\beta$-equilibrium limit of equation of state (EOS), constrained via a Bayesian inference methodology. The EOS is constrained by minimal approximations on a few nuclear saturation properties, low-density pure neutron matter constraints from chiral effective field theory, and a neutron star (NS) maximum mass greater than 2.0 $M_{\odot}$. Two sets of EOS derived from the relativistic mean field model for nucleonic and hyperonic matter constrained by a Bayesian inference calculation at the zero temperature limit are used. The thermal adiabatic index ($\Gamma_{\rm Th}$) is calculated as a function of the baryonic density across several temperatures for both the sets. Our results suggest that the maximum NS mass is of the order of 2.15 $M_\odot$ if hyperons are present. In addition, the present study suggests that an observation of NS with mass larger than $2.2\ M_{\odot}$ can indirectly indicates the absence of hyperons in its core. The deleptonization of hyperonic PNS reduces the stellar maximum mass rendering the PNS exceeding the zero temperature maximum stellar (baryonic) mass limit becomes metastable which is prone to collapse into a black hole while PNS below such a mass threshold evolves to a stable NS.
\end{abstract}

\maketitle

\section{Introduction}
\ac{pns} form as a consequence of successful supernova explosions when the stellar remnant is gravitationally decoupled from the expanding ejecta \cite{Burrows:1986me}. The neutrinos emitted by the remnant are crucial to supernova energetics and may also play an important role in supernova nucleosynthesis. These neutrinos, with their energies and emission timescales, provide valuable insights into the mass and composition of \ac{pns}. Determining the exact \ac{eos} governing a \ac{pns} is a complex and compelling topic in current astrophysical research and debate. The \ac{eos} of a \ac{pns} is influenced by three thermodynamic parameters, commonly chosen as the temperature $T$, the baryon number density $n_{\rm b}$, and the lepton fraction $Y_{\rm lep}=n_{\rm l}/n_{\rm b}$, with $n_l$ the lepton  number density. These parameters must cover wide ranges: $10^{-14}~{\rm fm}^{-3} \leq n_{\rm b} \leq 1.5~{\rm fm}^{-3}$, $0 \leq Y_{\rm lep} \leq 0.6$ and $0 \leq T \leq 100~{\rm MeV}$ \cite{Pons:1998mm, Sumiyoshi_2007, Janka_PhysRep_2007, Fischer_2009, Shibata_11, OConnor_2011, hempel12, SHF_ApJ_2013, Mezzacappa2015, Rosswog_15, Baiotti_2017, Connor2018ApJ, Burrows2020MNRAS, Ruiz2020, Kunkel:2024otq}.

As the star collapses, neutrinos are produced in large quantities through electron capture and are mostly temporarily prevented from escaping because their mean free path is small compared to the radius of the star. During this trapped neutrino era of the \ac{pns} evolution, the entropy per baryon is about 1 (with $k_B=1$) throughout the star and the total number of leptons per baryon is about $Y_e+Y_{\nu_e} \simeq 0.4$ in an initial state. The neutrinos in the core inhibit the possibility of having exotic matter like hyperonic matter or quark matter. As the \ac{pns} cools, the neutrino mean free path increases and the neutrinos leave the star on a timescale of 20--60 seconds. During such deleptonization, the neutrino diffusion heats up matter, and the entropy per baryon increases to 2. With deleptonization, it is possible to have hyperons at the core. In such a scenario, the leptonic content and the maximum mass decrease. \ac{pns} with baryonic masses above the \ac{tov} baryonic maximum mass \footnote{the maximum baryonic mass obtained for a stable neutron star from the integration of the \ac{tov} equations} for neutrino free cold matter are metastable and will collapse to a black hole before it completely cools down \cite{Prakash:1996xs}. If the mass of the \ac{pns} is below this mass threshold, the star will be stable and cool into a \ac{ns} as the neutrinos carry energy away from the star.

The various numerical studies of these diverse phenomena developed in the last decades have shown significant sensitivity to the \ac{eos}, see e.g. \cite{Pons:1998mm, Janka_2012, Bauswein_2012, Koeppel_2019, Bauswein_PRL_2020, Preau_MNRAS_2021, Kunkel:2024otq}. The \ac{eos} for \ac{pns} is critical for their evolution and neutrino emission. In \cite{Pons:1998mm}, the thermal and chemical evolution was studied during the Kelvin-Helmholtz phase of \ac{ns} birth using a neutrino opacity consistent with the \ac{eos}. The evolution of a \ac{pns} is characterized by the Kelvin-Helmholtz cooling phase, a period during which the star undergoes significant thermal and compositional changes. Initially, the \ac{pns} is hot and rich in neutrinos (lepton-rich). But as it emits neutrinos, it gradually becomes deleptonized and loses its thermal energy, transforming into a cold, neutrinoless, neutron rich star. The rate and nature of this cooling process are governed by the microphysical properties of dense matter within the \ac{pns} which is encapsulated in the \ac{eos}, and the opacity of matter for neutrinos that affects how efficiently neutrinos can escape from the \ac{pns} matter. \ac{pns} cooling simulations, including potential hyperons, explore how initial stellar models, total mass, \ac{eos} variations, and hyperons affect stellar evolution and detectable neutrino signals. In \cite{Raduta:2021coc}, the authors have evaluated the \ac{eos} models in the CompOSE database, emphasizing their applicability in numerical simulations of core-collapse supernovae, \ac{bns} mergers, and \ac{pns} evolution. They emphasize that purely nucleonic models, consistent with astrophysical and nuclear constraints, reveal significant influences of the nucleon effective mass on thermal properties in these extreme environments. Reviews of microscopic many-body techniques and phenomenological frameworks emphasize the importance of considering nucleons and hyperons, contrasting theoretical predictions with empirical observations \cite{Burgio:2021vgk}. 

In \cite{Raithel:2019gws} a general framework was proposed for the accurate calculation of the thermal pressure of \ac{npe} matter at any given density, temperature and proton fraction.  This quantity is essential for modeling astrophysical phenomena such as supernovae and \ac{bns} mergers. The method considered, which uses five physically motivated parameters, captures the leading-order effects of degenerate matter on thermal pressure, improving accuracy over existing models by 1-3 orders of magnitude.  This framework also allows the extension to finite temperature of cold, parametric and non-parametric \ac{eos}s, which do not include microphysics.  The effects of magnetic fields and rotation on \ac{pns}s have also been investigated in \cite{Franzon:2016iai} within a hadronic chiral SU(3) model, or in \cite{Rabhi:2009ii, Rabhi:2011ej} within a \ac{rmf} description. Strong poloidal magnetic fields were considered to significantly deform the stars and alter their structure, composition, and trapped neutrino populations, affecting the strangeness content and temperature of the stars throughout their evolution.

Neutron star matter, with hyperonic degrees of freedom and trapped neutrinos, is richer in neutrinos than nucleonic matter. The extra pressure of neutrinos allows \ac{pns} to support a larger gravitational and baryonic mass compared to neutrinoless cold \ac{ns}. As a consequence, deleptonization reduces the maximum gravitational and baryonic mass of the star, which is not compensated by thermal energy. This fact may cause the delayed decay of the \ac{pns} into a \ac{bh}. In this scenario, it is assumed that accretion occurs only during the initial phase immediately after the supernova, and the evolution of \ac{pns} occurs at a fixed baryonic mass \cite{Keil:1995hw, Prakash:1996xs, Vidana:2002rg}. A similar scenario could occur if a kaon condensate forms in the interior of \ac{pns} \cite{Brown:1993jz}.

The \ac{rmf} approach is a widely used theoretical framework to model the \ac{eos} of \ac{pns} \cite{Oertel:2016bki}. This approach incorporates interactions mediated by mesons (such as $\sigma$, $\omega$, $\rho$ mesons) to describe the behavior of nucleons at high density, and easily includes degrees of freedom other than nucleons and leptons, such as hyperons, kaons or deconfined quarks. 

Recent advances in understanding dense matter \ac{eos}, together with improvements in neutrino transport models, have provided a more detailed picture of the \ac{pns} cooling process \cite{Martinez-Pinedo:2012eaj, Nakazato:2019ojk, Reddy:1998hb}. This progress has been crucial for making more accurate predictions about the behavior of \ac{pns} and the signals they produce, especially for neutrino detectors. However, to fully understand and predict these phenomena, ongoing simulations and detailed calculations are required, highlighting the complex interplay of nuclear physics, thermodynamics, and astrophysics in the study of \ac{pns}. The \ac{eos} derived from an RMF model can predict various structural and thermal properties of \ac{pns}, such as their mass-radius relationship, stability and cooling behavior. It also helps to understand the role of neutrinos, which are trapped in the dense matter of a \ac{pns} and significantly influence its evolution. 

This work aims to improve the understanding of the finite temperature \ac{eos}, focusing on nucleonic and hyperonic models in the \ac{rmf} theory of dense matter. In this study, we use a \ac{rmf} model in which the $\beta$-equilibrated \ac{eos} at zero temperature is constrained by a Bayesian inference method subject to minimal constraints on a few nuclear saturation properties \cite{Kumar:2025efd}. The low-density pure neutron matter is constrained by chiral effective field theory ($\chi$\ac{eft}) and the high-density nuclear matter is constrained by astrophysical observations such as \ac{ns} mass. The \ac{eos} data set is evaluated including hyperons in the composition. This \ac{eos} set is used by switching the presence of the hyperonic content for different fixed temperatures ($T=0,\ 10,\ 30$ and $50$ MeV), entropies per baryon ($S=0,\ 1,$ and $2$) and lepton fractions ($Y_l = 0.4$ or neutrinoless $\beta$-equilibrium). Our analysis includes results for the thermal adiabatic index $\Gamma_{\rm Th}$, which can be used to account for the thermal effects on the zero temperature \ac{eos}. Old \ac{ns}s have temperatures below 1 MeV, while \ac{pns}s can reach central temperatures of around 50 MeV or even higher \cite{Burrows:1986me, Pons:1998mm, Kunkel:2024otq}. In the context of compact star mergers, temperatures can reach up to 80 MeV \cite{Galeazzi:2013mia}. These temperature levels are comparable to those observed in heavy-ion collision experiments at \ac{rhic} and \ac{lhc} \cite{Schenke:2011tv, Alqahtani:2017jwl}. Therefore, given these temperatures, it is logical to use the same mathematical models or similar approaches to describe these systems.

The structure of the article is as follows: Sec. \ref{sec:formalism} covers the formalism of temperature dependent \ac{eos} models. In Sec. \ref{sec:results_and_discussions} we present the results and findings of the present study. Finally, \ref{sec:summary_and_conclusion} provides a concluding summary, highlighting the key points of our research. Throughout this paper we use natural units with $\hbar = c = G = k_B = 1$.

\section{Formalism} \label{sec:formalism}
In the present study, we explore a relativistic framework of the temperature dependent \ac{eos} of dense matter in the core of \ac{pns}s. We employ the \ac{rmf} model with non-linear mesonic interactions. Further, we examine the influence of the inclusion of baryons with strangeness, e.g. hyperons ($\Lambda$, $\Sigma^{+,-,0}$, $\Xi^{0,-}$), on the \ac{eos}. The interactions among baryons are facilitated by the exchange of the isoscalar-scalar ($\sigma$), the isoscalar-vector ($\omega$) and  the isovector-vector ($\varrho$) mesons. For the models including hyperons, we also include other mesons with strange degrees of freedom - the hidden strangeness isoscalar-vector ($\phi$), the isoscalar-scalar ($\sigma^*$) meson. The Lagrangian describing the baryonic degrees of freedom is expressed as follows:
\begin{equation}
\mathcal{L}=  \sum_{\rm b} \mathcal{L}_{\rm b} + \mathcal{L}_{\rm M} + \mathcal{L}_{\rm int}, \label{lagrangian}
\end{equation} 
with
\begin{eqnarray}%{rCl}
\mathcal{L}_{\rm b} &=& \bar{\Psi}_{\rm b}\big[\gamma^{\mu}\left(i \partial_{\mu}-g_{\omega {\rm b}} \omega_{\mu} - g_{\varrho {\rm b}} {\boldsymbol{t}} \cdot \boldsymbol{\varrho}_{\mu} - g_{\phi {\rm b}} \phi_{\mu}\right) \nonumber \\
&& \qquad \qquad - \left(m - g_{\sigma {\rm b}}\sigma- g_{\sigma^* {\rm b}}\sigma^*\right)\big] \Psi_{\rm b}, 
% \end{IEEEeqnarray}
\\
&& \nonumber
\\
% \begin{eqnarray}{rCl}
\mathcal{L}_{\rm M} &=& \frac{1}{2}\left[\partial_{\mu} \sigma \partial^{\mu} \sigma-m_{\sigma}^{2} \sigma^{2} \right] + \frac{1}{2}\left[\partial_{\mu} \sigma^* \partial^{\mu} \sigma^*-m_{\sigma^*}^{2} \sigma^{*2} \right] \nonumber \\
&& - \frac{1}{4} \boldsymbol{F}_{\mu \nu}^{(\varrho)} \cdot \boldsymbol{F}^{(\varrho) \mu \nu} + \frac{1}{2} m_{\varrho}^{2} \boldsymbol{\varrho}_{\mu} \cdot \boldsymbol{\varrho}^{\mu} \nonumber \\
&& - \frac{1}{4} \sum_{O=\omega,\phi} \left[ F_{\mu \nu}^{(O)} F^{(O) \mu \nu} + 2m_{O}^{2} O_{\mu} O^{\mu} \right],
% \end{eqnarray}
\\
&& \nonumber
\\
% \begin{IEEEeqnarray}{rCl}
\mathcal{L}_{\rm int} &=& - \frac{1}{3} b g_\sigma^3 \sigma^{3}-\frac{1}{4} c g_\sigma^4 \sigma^{4}+\frac{\xi}{4!}(g_{\omega}\omega_{\mu}\omega^{\mu})^{4} \nonumber \\
&& + \Lambda_{\omega}g_{\varrho}^{2}\boldsymbol{\varrho}_{\mu} \cdot \boldsymbol{\varrho}^{\mu} g_{\omega}^{2}\omega_{\mu}\omega^{\mu}.
\end{eqnarray}

The field $\Psi_{\rm b}$ is a Dirac spinor that describes baryons with a bare mass $m_{\rm b}$. The $\gamma^\mu $  are the Dirac matrices and $\boldsymbol{t}$ is the isospin operator. The vector meson field tensors are defined as $F^{(\omega, \varrho, \phi)\mu \nu} = \partial^{\mu} A^{(\omega, \varrho, \phi)\nu} -\partial^{\nu} A^{(\omega, \varrho, \phi) \mu}$ (the non-linear term of the $\rho$ meson tensor does not contribute in the mean-field approximation). The $g_{\sigma}$, $g_{\sigma^{\star}}$, $g_{\omega}$, $g_{\varrho}$, and $g_{\phi}$ are the couplings of the baryons to the meson fields $\sigma$,$\sigma^*$, $\omega$,  $\varrho$, and $\phi$ of masses $m_\sigma$, $m_{\sigma^{\star}}$, $m_\omega$,  $m_\varrho$, and $m_\phi$ respectively. 

The parameters $b$, $c$, $\xi$ and $\Lambda_{\omega}$, which define the strength of the non-linear terms, are determined together with the couplings $g_i$ ($i=\sigma,\, \omega,\,\varrho$), imposing a set of constraints.  The terms with $b$, and $c$, have been introduced in \cite{Boguta:1977xi} to control the nuclear matter incompressibility at nuclear saturation density. The $\xi$ term controls the stiffness of the \ac{eos} at high densities, the larger its value  the softer the \ac{eos}  at high densities. The $\Lambda_{\omega}$ parameter affects the density dependence of the symmetry energy. Increasing the $\Lambda_{\omega}$ parameter implies a decrease in the slope of the symmetry energy at saturation. The effect of the non-linear terms of the meson fields is clearly seen from the equations of motion for the mesons
\begin{IEEEeqnarray}{rCl}
m_{\sigma}^{2}{\sigma} + {b g_\sigma^3}{\sigma}^{2} + {c g_\sigma^4}{\sigma}^{3} &=& g_{\sigma} \sum_{\rm b} x_{s {\rm b}} \rho^s_{\rm b} \label{sigma} 
\\
% \begin{IEEEeqnarray}{rCl}
m_{\omega}^{2}{\omega} + \frac{\xi}{3!}g_{\omega}^{4}{\omega}^{3} + 2\Lambda_{\omega}g_{\varrho}^{2}g_{\omega}^{2}{\varrho}^{2} \omega &=& g_{\omega} \sum_{\rm b} x_{\omega {\rm b}}\rho_{\rm b} \label{omega}
% \end{IEEEeqnarray}
\\
% \begin{IEEEeqnarray}{rCl}
m_{\varrho}^{2}{\varrho} + 2\Lambda_{\omega}g_{\omega}^{2}g_{\varrho}^{2}{\omega}^{2}\varrho &=& g_{\varrho} \sum_{\rm b} x_{\rho {\rm B}} I_{3} \rho_{\rm b}, \label{rho} 
% \end{IEEEeqnarray}
\\
m_{\sigma^*}^{2}{\sigma^*}  &=& g_{\sigma^*} \sum_{\rm b} x_{s^* {\rm b}} \rho^s_{\rm b} \label{sigma*} \\
% \begin{IEEEeqnarray}{rCl}
m_{\phi}^{2}{\phi} &=& g_{\phi} \sum_{\rm b} x_{\phi {\rm B}} \rho_{\rm b}, \label{phi}
\end{IEEEeqnarray}
where $\rho_{\rm b}^s = <\bar\psi_{\rm b}\psi_{\rm b}>$ and $\rho_{\rm b} = <\bar\psi_{\rm b}\gamma_{0} \psi_{\rm b}>$ are the scalar density and the vector density of baryon $b$, respectively. The $\sigma$, $\sigma^*$,  $\omega$, $\varrho$ and $\phi$ designate here the mean field values of the scalar fields $\sigma$, $\sigma^*$ and the time-like components of the vector fields $\omega$, $\varrho$, $\phi$, where the space-like components are zero. The coupling ratios of meson $j$ with respect to baryon $i$ are $x_{ji} = g_{ji}/g_j$ for $j = \sigma,\, \sigma^{\star},\, \omega,\, \rho,\, \phi$, taking $x_{jN}=1, \, j=\sigma,\,\omega,\, \varrho$, $x_{jN}=0, \, j=\sigma^*,\phi$ and $g_{\phi} = g_{\omega}$, $g_{\sigma^*} = g_{\sigma}$, see \cite{Fortin_PRC_2017}.

The baryon $b$'s vector and scalar number densities are defined as
\begin{IEEEeqnarray}{rCl}
n_{\rm b} &=& 2 \int \frac{{\rm d}^3k}{(2\pi)^3} \big( f_{{\rm b}+} - f_{{\rm b}-}\big), \label{vector_density} \\
n_{\rm b}^s &=& 2 \int \frac{{\rm d}^3 k}{(2\pi)^3} \frac{m_{\rm b}^*}{E_{\rm b}^*} \big(f_{{\rm b}+} + f_{{\rm b}-}\big). \label{scalar_density}
\end{IEEEeqnarray}
with the distribution functions $f_{{\rm b}\pm}$ for particles and antiparticles defined as
\begin{equation}
f_{{\rm b}+}={\frac{1}{e^{\beta(E_{\rm b}^* - \tilde{\mu}_{\rm b})} + 1}}, \quad f_{{\rm b}-} = {\frac{1}{e^{\beta(E_{\rm b}^* + \tilde{\mu}_{\rm b})} + 1}} \label{fb}
\end{equation}
respectively, where $\beta = 1/T$ with $T$ as the temperature, and $E_b^* = \sqrt{k^2 + m_{\rm b}^{*2}}$. Here the effective baryon mass and effective chemical potentials are defined as follows
\begin{IEEEeqnarray}{rCl}
m^{*}_{\rm b} &=& m_{\rm b} - g_{\sigma {\rm b}} \sigma- g_{\sigma^* {\rm b}} \sigma^*, \label{effective_mass} \\
\tilde{\mu}_{\rm b} &=& \mu_{\rm b} - g_{\omega {\rm b}} \omega - g_{\rho {\rm b}} t_{3{\rm b}} \varrho - g_{\phi {\rm b}}\phi. \label{effective_chemical_potential}
\end{IEEEeqnarray}
The energy density and entropy density within the model, Eq (\ref{lagrangian}) at various temperature are given as
\begin{IEEEeqnarray}{rCl}
\epsilon &=& \sum_{\rm b} \epsilon_{\rm kin, b}(T) \nonumber \\
&& + \frac{1}{2} \big( m_{\sigma}^{2}{\sigma}^{2} + m_{\sigma^*}^{2}{\sigma^*}^{2} + m_{\omega}^{2}{\omega}^{2} + m_{\varrho}^{2}{\varrho}^{2} + m_{\phi}^{2}{\phi}^{2} \big) \nonumber \\
&& + \frac{b}{3}(g_{\sigma}{\sigma})^{3} +\frac{c}{4}(g_{\sigma}{\sigma})^{4}+\frac{\xi}{8}(g_{\omega}{\omega})^{4} +3 \Lambda_{\omega}(g_{\varrho}g_{\omega}{\varrho}{\omega})^{2}, \nonumber \\
&& \label{energy_density}
% \end{IEEEeqnarray}
\\
% \begin{IEEEeqnarray}{rCl}
s &=& - \frac{2}{(2\pi)^3} \sum_{\rm b} \int {\rm d}^3k \Big[ f_{\rm b-} \ln f_{\rm b-} + (1 - f_{\rm b-}) \times \nonumber \\
&& \qquad \qquad \qquad \qquad \ln (1 - f_{\rm b-}) + \left(- \to + \right)\Big], \label{entropy_density}
\end{IEEEeqnarray}

\noindent and the pressure is determined from the thermodynamic relation as follows
\begin{equation}
p = \sum_{\rm b}\mu_{\rm b}\rho_{\rm b} - T s_{\rm } - \epsilon. \label{pressure}
\end{equation}

\noindent where the kinetic part of the energy density ($\epsilon_{\rm kin, B}(T)$) is given as
\begin{IEEEeqnarray}{rCl}
\epsilon_{\rm kin, b}(T) = 2 \sum_{\rm b} \int \frac{{\rm d}^3 k}{(2\pi)^3} E_{\rm b}^* \big( f_{\rm b+} + f_{\rm b-}\big) \nonumber
\end{IEEEeqnarray}

At zero temperature, the distribution functions defined in Eq. (\ref{fb}) reduce to step functions and the integrals extend up to the Fermi surface with Fermi momentum $k_{\rm Fb}$. We can define the scalar, vector densities in zero temperature limit as
\begin{IEEEeqnarray}{rCl}
n_{\rm b} &=& \frac{k_{\rm Fb}^3}{3\pi^2}, \\
n_{\rm b}^s &=& 2 \int \frac{{\rm d}^3 k}{(2\pi)^3} \frac{m_{\rm b}^*}{E_{\rm b}^*}.
\end{IEEEeqnarray}

The kinetic energy density and pressure, in the zero temperature limit, is given by
\begin{IEEEeqnarray}{rCl}
\epsilon_{\rm kin, b}(T = 0) &=& 2 \int_0^{k_{\rm Fb}} \frac{{\rm d}^3k}{(2\pi)^3} \sqrt{k^2 + {m_{\rm b}^*}^2} \\
p_{\rm kin, b}(T = 0) &=& \frac{2}{3} \int_0^{k_{\rm Fb}} \frac{{\rm d}^3k}{(2\pi)^3} \frac{k^2}{\sqrt{k^2 + {m_{\rm b}^*}^2}}.
\end{IEEEeqnarray}
We consider electrons ($e$) and electron-neutrino ($\nu_e$) as the lepton contribution to matter in the \ac{pns}. To determine the complete \ac{eos} of the system, the corresponding lepton energy density, entropy density and pressure are added to the baryons contribution.

In the context of \ac{pns} evolution, the dynamical time scales are much longer than the weak interaction time scales so that the matter is in $\beta$-equilibrium which in turn imply various chemical potentials satisfy the relations
\begin{equation}
\mu_{\rm b} = \mu_n - q_{\rm b}(\mu_e - \mu_{\nu_e}),
\end{equation}
where, $q_{\rm b}$ is the electrical charge of $b$th baryon. 

We will consider the thermodynamics for two conditions relevant to \ac{pns} evolution. The first scenario  corresponds to neutrinos trapped in the stellar matter, an entropy per baryon $S = s/\rho_{\rm b}\simeq 1$ and the electron lepton number $Y_{\rm lep} ={(\rho_e + \rho_{\nu_e})}/{\rho_{\rm B}} \simeq 0.4$ \cite{Prakash:1996xs}, i.e. the concentration of leptons per baryon is of the order of $0.4$. The number of leptons is conserved here. In a later stage of evolution, the neutrinos completely flow out of the \ac{pns} and the lepton number is no longer conserved. Only the baryon charge and the electric charge remain conserved. Without lepton number conservation, $\mu_\nu=0$, the neutrinos fall out of the $\beta$-equilibrium and we have
\begin{equation}
\mu_{\rm b} = \mu_n - q_{\rm b} \mu_e.
\end{equation}
Thus, while the compositional properties of a cold, mature neutron star depend on a single parameter \ac{eos} relating pressure to energy density, the \ac{eos} for \ac{pns} depends on additional parameters such as a fixed lepton fraction ($Y_L$) and a fixed entropy per baryon.

Many numerical simulations with finite temperature \ac{eos} have  been done  starting from a cold \ac{eos} with a phenomenological extension to finite temperatures using an ideal gas like behavior for the thermal contribution to the pressure  written as
\begin{equation}
p = p_{\rm cold}+(\Gamma_{\rm Th}-1)\epsilon_{\rm Th}
\end{equation}
where the thermal adiabatic index $\Gamma_{\rm Th}$, an important quantity for the supernovae simulations and \ac{bns} merger simulations, takes a value in the range
$1.5\le\Gamma_{\rm Th}\le 2$ and is assumed to be constant for all pressures $p$ and energy densities $\epsilon$. The thermal index for a given \ac{eos} is
\begin{equation}
\Gamma_{\rm Th}=1+\frac{p_{\rm Th}}{\epsilon_{\rm Th}}
\label{gth}
\end{equation}
where $p_{\rm Th} = p(T) - p(T=0)$ and $\epsilon_{\rm Th} = \epsilon(T) - \epsilon(T=0)$. We will compare the above result with the range of values $\Gamma_{\rm Th}$ usually takes. We will also discuss the validity of the $\Gamma$ law approximation of thermal effects for the different scenarios considered, in particular, fixed density, temperature, or lepton fraction.

\section{Results and Discussion} \label{sec:results_and_discussions}
In the following, we analyse two sets of 18000 EOSs each at finite temperature, one considering only nucleonic matter and a second including also hyperons. The sets have been derived from a Bayesian inference calculation at zero temperature \cite{Kumar:2025efd}, subject to some minimal nuclear matter constraints, including $\epsilon_0$, the binding energy per nucleon; $K_0$, the incompressibility; and $J_{\rm sym,0}$, the symmetry energy at nuclear saturation density $\rho_0$. In addition, the low density pure neutron matter \ac{eos} from an N3LO calculation in $\chi$\ac{eft}, and the astrophysical observation of a \ac{ns} mass exceeding 2$M_\odot$, are imposed \cite{Malik:2023mnx}. As exotic degrees of freedom,  we consider the hyperons $\Lambda$, $\Sigma^{\pm 0}$ and $\Xi^{0-}$.  At $T=0$, $\Lambda$ is usually the first hyperon to appear, being the lightest, partly because the $\Sigma$ hyperon couples repulsively to nuclear matter, as the nonexistence of $\Sigma$ hypernuclei indicates \cite{Gal:2016boi}, although its negative charge would favor its presence in order to replace electrons.  The onset of $\Sigma^-$ is often delayed to densities greater than the onset of $\Xi^-$ \cite{Weissenborn:2011kb, Fortin_PRC_2016, Fortin:2020qin, Stone:2019blq} and is generally absent within stable \ac{ns}. However, at finite temperature the role of the interaction weakens with increasing temperature, and the magnitude of the mass becomes more important in defining the particle abundances, with the fractions of $\Sigma$ becoming dominant over those of $\Xi$ \cite{Oertel:2016bki, Fortin:2017dsj}.

We use these two sets of \ac{eos}s with and without hyperons considering: i) a fixed temperature ($T=0,\ 10,\ 30$ and $50$ MeV); ii) a fixed entropy per baryon ($S=0,\ 1$ and $2$), for $\beta$-equilibrated matter, to determine \ac{eos}s and its properties applicable to \ac{ns}s. We consider a fixed lepton fraction scenario ($Y_{\rm lep} = 0.4$) with entropy per baryon $S=1$, corresponding to the trapped neutrino scenario, and a scenario with $S=2$ and $Y_{\nu_e}=0$, corresponding to a deleptonisation stage of the \ac{pns} evolution.

\begin{figure*}
    \centering
    \includegraphics[width=1.0\linewidth]{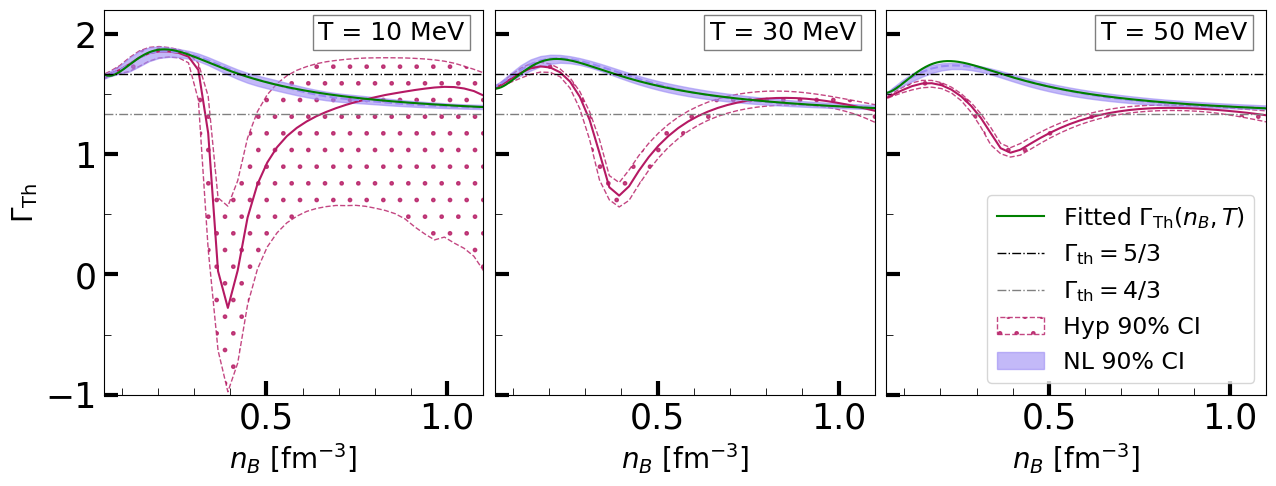}
    \caption{The thermal adiabatic index, denoted as $\Gamma_{\rm Th} = 1 + \dfrac{p_{\rm Th}}{\epsilon_{\rm Th}}$, where $\epsilon_{\rm Th} = \epsilon(T) - \epsilon(0)$ and $p_{\rm Th} = p(T) - p(0)$, is computed for the hyperon set (shown in dotted rose) and the nucleonic set (shown in blue) at constant temperatures of $T = 10,\, 30,$ and $50$ MeV, respectively, shown across the panels from left to right.  We additionally provide an fitted $\Gamma_{\rm Th} (n_B, T)$ curve specifically for the nucleonic scenario (refer to the text for further details).}
    \label{fig:gamma}
\end{figure*}
\begin{figure*}
    \centering
    \includegraphics[width=0.85\linewidth]{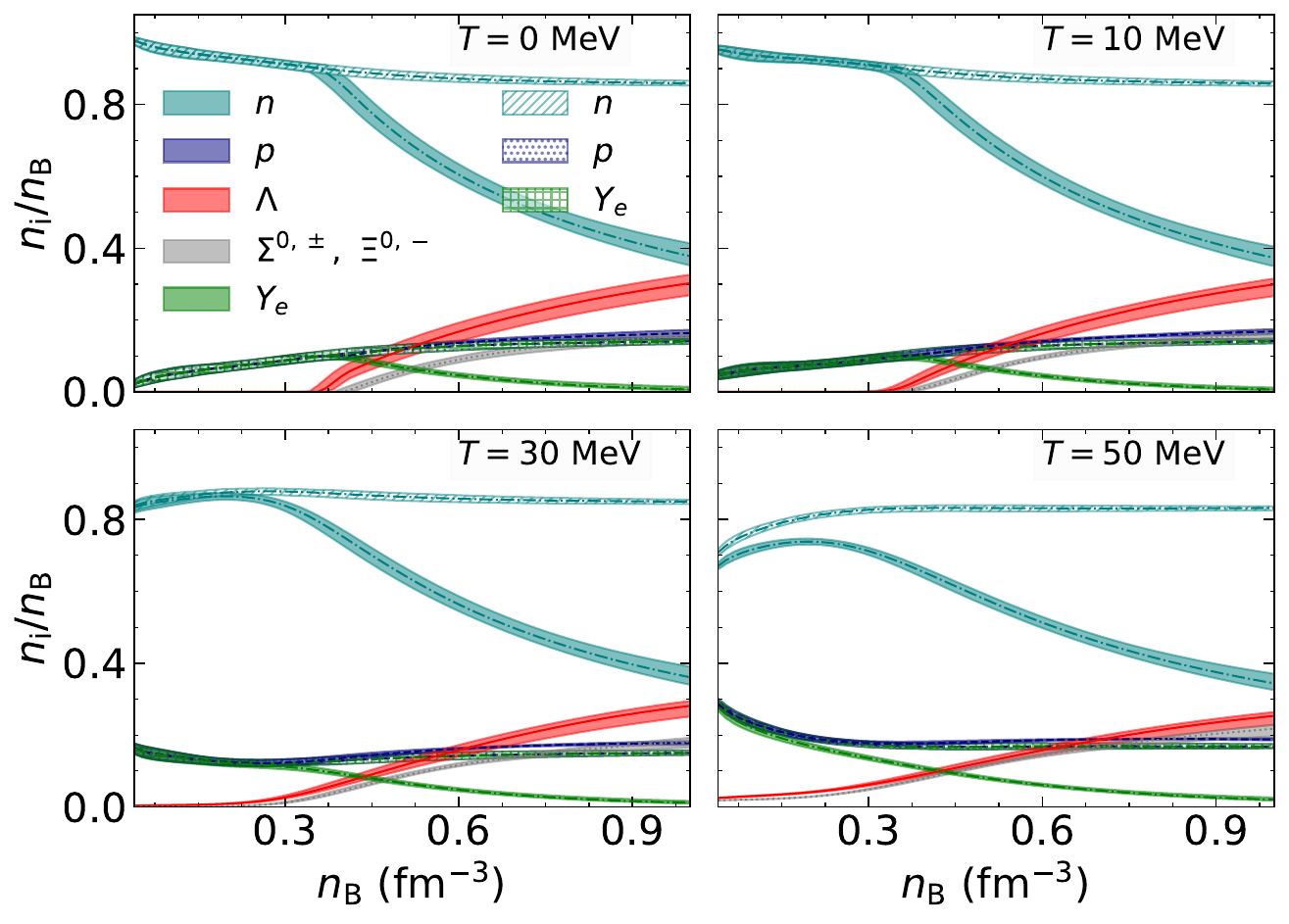}
    \caption{The $90\%$ CL of the particle fractions as a function of total baryon density for different temperatures, ($T = 0,\,10,\,30,\,50\ {\rm MeV}$). The different colors denote (i) $n$ neutron (green), (ii) $p$ proton (salmon), (iii) $\Lambda$ hyperon (pink), and all other hyperons like (iv) $\Sigma$ hyperon, (v) $\Xi$ hyperon (silk blue). The hatched curves are corresponding to the $npe$ matter.} 
    \label{fig:hyp_particle_fraction_final}
\end{figure*}

In Fig. \ref{fig:gamma}, we show the thermal adiabatic index, $\Gamma_{\rm th}$, as defined in Eq. (\ref{gth}). This index is calculated for the hyperonic set (in rose dotted pattern) and the nucleonic set (in blue), the medians are presented respectively, by a full rose line and a dashed blue line, and the demarked bands correspond to 90\% confidence intervals (CI)s. The figure consists of three panels for different temperatures: 10 MeV, 30 MeV, and 50 MeV. At 10 MeV, the hyperon set shows significant fluctuations in $\Gamma_{\rm Th}$ across densities from 0.1 to 1.0 fm\(^{-3}\), while the nucleonic set shows a smoother behavior. Similar results were obtained in Ref. \cite{Kochankovski:2022rid} within the FSU2H model and in Ref. \cite{Raduta:2021coc} for the models in the CompOSE database. The fast decrease of $\Gamma_{\rm th}$ is related to the onset of hyperons: the thermal energy is distributed by a larger number of degrees of freedom and the pressure suffers a strong softening. The increase of degrees of freedom occurs at $\sim 0.4$ fm$^{-3}$ for $T=0$ and 10 MeV, as  can be seen in Fig. \ref{fig:hyp_particle_fraction_final} where the particle abundances are displayed as a function of density for different temperatures. At 30 MeV, the hyperons variations are smaller but still display notable features, such as a dip around 0.5 fm\(^{-3}\), whereas the nucleonic set shows a smooth behavior with a smaller amplitude between minimum and maximum values. The different behavior obtained for matter with hyperons reflects the fact that hyperons set in at much smaller densities and contribute from very low densities, although the strongest contribution still occurs close to the onset densities at $T=0$. At 50 MeV, both sets exhibit smoother curves, with the hyperon set variations becoming less pronounced, and the nucleonic set maintaining a relatively steady trend. Notice that as the temperature increases the bands defining the thermal index with and without hyperons become narrower indicating that the thermal effects and not the interaction control the behavior. The width of the bands in the left panel reflect the differences between the dataset \ac{eos} due to the different parameterizations of the interaction of mesons with nucleons. Including hyperons broadens the interaction differences. At large temperature the potential contribution becomes much smaller than the kinetic contribution to pressure and energy density reducing the originally wide bands to almost a line. Along with the interacting nucleonic and hyperonic matter system, we also plot the constraints values of $\Gamma_{\rm Th} = \frac{4}{3} \left( \frac{5}{3}\right)$ for a dilute ideal ultra-relativistic (non-relativistic) Fermion gas for comparison. As may be observed at small densities, $\Gamma_{\rm Th}$ starts with the non-relativistic value ($\Gamma_{\rm Th} = \frac{4}{3}$) while at high densities it approaches its ultra-relativistic limit ($\Gamma_{\rm Th} = \frac{5}{3}$) as expected. We also have obtained a parametric expression for $\Gamma_{\rm th}$ as a function of baryon density ($n_B$) and temperature (T) given as 

\begin{IEEEeqnarray}{rCl}
\Gamma_{\rm Th}^{\rm nuc}(n_B, T) = a + b \cdot c^{n_B} \cdot \rho^{T^{d}} \label{gth}
\end{IEEEeqnarray}

where $n_B$ and $T$ are written in units of fm$^{-3}$ and MeV, respectively. The other fitted coefficients are given as  $a=1.3665$, $b=11.9638$, $c=0.0022$ and $d=0.0867$. We have checked that such a parametrization is valid for density $n_B \in [0.08, 1.2]$ fm$^{-3}$, and temperature $T \in [5,100]$ MeV. We also mention here that such a parametrization while gives a reasonable descriptions for nucleonic matter but dose not give a good description for hyperonic matter. This may be expected as  $\Gamma_{\rm Th}$ for hyperonic matter shows more stature as compare to nucleonic case. Such a relation could be useful for binary neutron star merger simulations.

In Fig. \ref{fig:hyp_particle_fraction_final} we present the particle fractions (90\% CL) at different densities, for different temperatures in each panel. In each band, the middle curve represents the median value of the particle fraction at different values of baryon density. We plot the particle fractions at temperature $T=0, 10, 30, 50\ {\rm MeV}$ from top-left, top-right, bottom-left, bottom-right respectively. The patterned distributions are plotted for \ac{npe} matter while homogeneous bands are for hyperonic matter. Light teal, light navy, and light red regions with various patterns show the fractions of neutron ($n$), proton ($p$), and lepton ($e$) in \ac{npe} matter while light teal, light navy, light red, light grey and light green regions without patterns are showing the fractions neutron ($n$), proton ($p$), lambda hyperon ($\Lambda$), the sum of other hyperons ($\Sigma^{0,\pm},\ \Xi^{0,-}$) and the sum of leptons ($e,\ \mu$) in hyperonic matter. As density increases, new particles emerge when the effective chemical potential of a particle becomes large enough at various temperatures ($T=0, 10, 30, 50\ {\rm MeV}$). For finite temperature, there is always a finite probability for the hyperon to set in for an effective chemical potential smaller than the effective mass, due to the thermal excitations of baryons. The threshold for the appearance of hyperons at $T=0$ depends upon the hyperon-nucleon potential. The inclusion of hyperons lowers the neutron abundances. At high enough densities the neutrons are highly energetic and can be replaced by the lambda hyperon ($\Lambda$) which already exceeds the proton abundance around three times the nuclear saturation density.  Such a behavior is due to the hyperon-nucleon potential and was also seen in earlier studies (see \cite{Weissenborn:2011kb,Bhowmick:2014pma,Kochankovski:2022rid}). Notice that for hyperonic matter as the temperature increases, the fractions of the different hadrons get closer and closer at large densities. The difference between neutrons and protons also decreases as $T$ increases. These behaviors reflect the increasing dominance of the kinetic contributions to the total energy.

\begin{figure}
    \centering
    \includegraphics[width=0.95\linewidth]{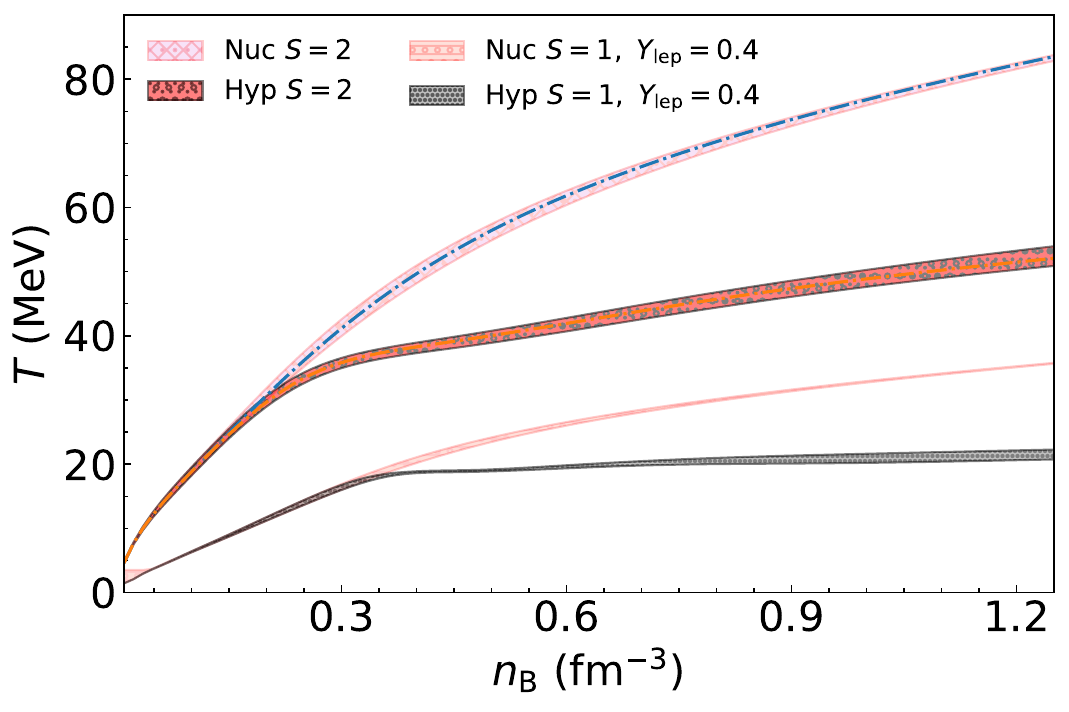}
    \caption{The temperature  ($90\%$ CI) as a function of the baryon number density with and without hyperons in \ac{pns}s for different values of entropy densities, (i) $S=1,\ Y_{\rm lep} = 0.4$, and (ii) $S=2$ for neutrino free matter in $\beta$-equilibrium.}
    \label{fig:tmp_fill}
\end{figure}

In Fig. \ref{fig:tmp_fill}, the temperature is given as a function of the density for four different scenarios: (i) $S=1,\ Y_{\rm lep}=0.4$, (ii) $S=2,\ Y_{\nu_e}=0$ for nucleonic matter and other two's for the hyperonic matter. It is quite impressive that the datasets give rise to such narrow bands at 90\% CI. As discussed in the previous studies \cite{Steiner:2000bi, Wei:2021veo, Rabhi:2011ej}, it is seen that the temperature is quite sensitive to the number of species. In this case, the larger the number of degrees of freedom the smaller the rise of the temperature with density. The temperature for an almost degenerate Fermi system (at low temperature) can be related with the entropy per baryon through \cite{Steiner:2000bi} as
\begin{IEEEeqnarray}{rCl}
T\sim \frac{S}{\pi^2}\frac{\sum_i k_{Fi}^3}{\sum_i k_{Fi}\sqrt{k_{Fi}^2+m_i^{*2}}}, 
\end{IEEEeqnarray}
such that, if the energy is shared by a larger number of particles, the Fermi momenta of each particle will be smaller (strongly affecting the numerator) and the mass term in the denominator will have a stronger effect. 
The difference between the two distributions obtained for nucleonic matter with $S=1$ is also due to the number of degrees of freedom: matter with $Y_l=0.4$ is more symmetric, the proton fraction is of the order of 0.3, and in addition to nucleons and electrons, neutrinos are also present. This leads to a lower temperature for fixed entropy in matter with trapped neutrinos compared to neutrino free matter. It is important to note that the data set including hyperons is quite restrictive, because the combination of including hyperons and imposing the lower gravitational maximum mass of two solar masses severely limits the parameter space.

\begin{figure}
    \centering
    \begin{tabular}{c}
    \includegraphics[width=0.99\linewidth]{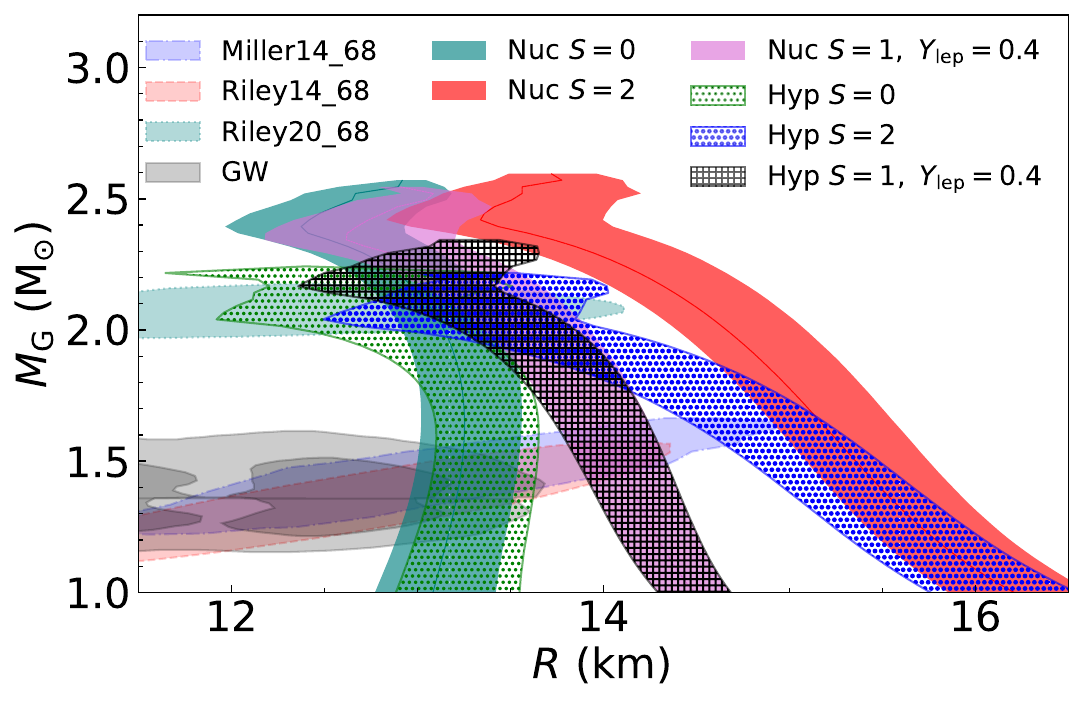}\\
    \includegraphics[width=0.99\linewidth]{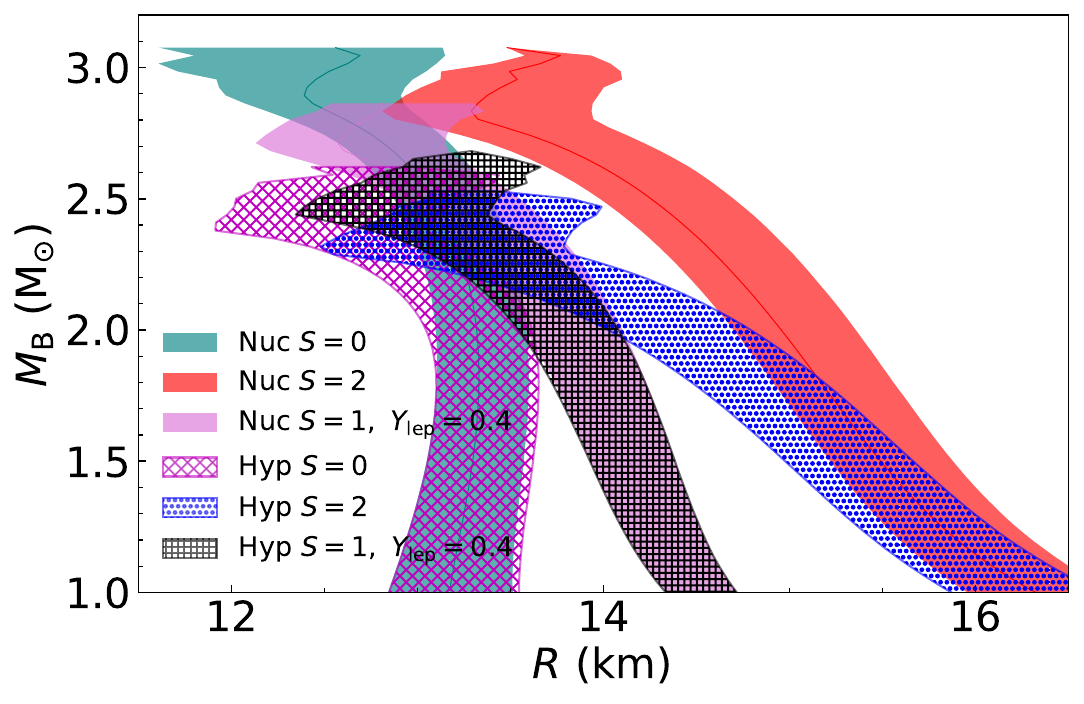}
    \end{tabular}
    \caption{The $90\%$ CI of mass-radius regions for different entropy per baryons, (i) $S=1,Y_{\rm lep}=0.4$ with trapped neutrinos and (ii) $S=2$ for neutrino free matter, with and without hyperons in \ac{pns}s. (Upper) the gravitational mass of a star while (Lower) baryonic mass. The light blue, dark blue, gold and gray coloured patches are representing the NICER and gravitational observations for the cold \ac{ns}s.  \label{fig:amr_am_and_amr_bm}}
\end{figure}

\begin{table*}[]
\caption{Various Neutron Star (\ac{ns}) properties for Nucleonic (Nuc) and Hyperonic (Hyp) sets at different entropy per baryon  ($S=0, 1, 2$) for  matter in $\beta$-equilibrium.}
\setlength{\tabcolsep}{2.5pt}
\renewcommand{\arraystretch}{1.2}
\begin{tabular}{|l|c|cc|c|cc|c|cc|c|cc|c|cc|c|cc|}
\toprule
\multirow{4}{*}{Quantity} & \multicolumn{6}{c|}{$S=0$, $Y_{\nu_e}=0$}   & \multicolumn{6}{c|}{$S=1$, $Y_{\rm lep}=0.4$}  & \multicolumn{6}{c|}{$S=2$, $Y_{\nu_e}=0$} \\ \cline{2-19} 
 & \multicolumn{3}{c}{Nuc} & \multicolumn{3}{|c|}{Hyp} & \multicolumn{3}{c}{Nuc} & \multicolumn{3}{|c|}{Hyp} & \multicolumn{3}{c}{Nuc} & \multicolumn{3}{|c|}{Hyp}                                \\ \cline{2-4} \cline{5-7} \cline{8-10} \cline{11-13} \cline{14-16} \cline{17-19} 
 & \multirow{2}{*}{Med} & \multicolumn{2}{c|}{90 \% CI} & \multirow{2}{*}{Med} & \multicolumn{2}{c|}{90 \% CI} & \multirow{2}{*}{Med} & \multicolumn{2}{c|}{90 \% CI} & \multirow{2}{*}{Med} & \multicolumn{2}{c|}{90 \% CI} & \multirow{2}{*}{Med} & \multicolumn{2}{c|}{90 \% CI} & \multirow{2}{*}{Med} & \multicolumn{2}{c|}{90 \% CI} \\ \cline{3-4} \cline{6-7} \cline{9-10} \cline{12-13} \cline{15-16} \cline{18-19} 
 &  & Min & Max &  & Min & Max & & Min & Max & & Min & Max & & Min & Max & & Min & Max \\ \hline
\hline
$M_{\rm max}$ & 2.394 & 2.348 & 2.458& 2.024 & 2.002 & 2.083& 2.403 & 2.360 & 2.466& 2.058 & 2.032 & 2.116& 2.424 & 2.387 & 2.485& 2.050 & 2.024 & 2.106 \\
$R_{\rm max}$ & 11.91 & 11.69 & 12.27& 11.78 & 11.54 & 12.16& 12.02 & 11.80 & 12.37& 11.73 & 11.49 & 12.12& 12.72 & 12.50 & 13.16& 12.33 & 12.08 & 12.75 \\
 $\epsilon_{\rm c,max}$ & 1072 & 1005 & 1121& 1116 & 1035 & 1165& 1048 & 990 & 1093& 1134 & 1051 & 1183& 1002 & 937 & 1040& 1125 & 1043 & 1172 \\
$p_{\rm c, max}$ & 473 & 393 & 527& 320 & 272 & 361& 460 & 381 & 508& 341 & 290 & 378& 420 & 348 & 461& 334 & 287 & 367 \\
\hline
$R_{1.2}$& 13.12 & 12.85 & 13.43& 13.12 & 12.85 & 13.43& 13.20 & 12.96 & 13.50& 13.19 & 12.95 & 13.49& 15.76 & 15.45 & 16.17& 15.60 & 15.29 & 16.02 \\
$R_{1.4}$& 13.18 & 12.93 & 13.48& 13.18 & 12.94 & 13.48& 13.28 & 13.06 & 13.58& 13.25 & 13.02 & 13.56& 15.44 & 15.16 & 15.83& 15.20 & 14.91 & 15.61 \\
$R_{1.8}$& 13.20 & 12.97 & 13.51& 13.10 & 12.85 & 13.45& 13.32 & 13.10 & 13.64& 13.11 & 12.86 & 13.48& 14.90 & 14.64 & 15.30& 14.28 & 13.95 & 14.74 \\
$R_{2.0}$& 13.10 & 12.88 & 13.44& 12.36 & 11.82 & 13.01& 13.23 & 13.01 & 13.57& 12.56 & 12.19 & 13.09& 14.58 & 14.32 & 15.01& 13.29 & 12.81 & 13.95 \\
\hline
\hline
\end{tabular}
\label{tab:s0_s1_s2}%
\end{table*}

\begin{table}
\centering
\caption{Maximum gravitational mass ($M_G$) and baryonic mass ($M_B$) for different equation of state models.}
\begin{tabular}{lcc}
\hline \hline 
\textbf{Case} & \textbf{Max. $M_G$ ($M_\odot$)} & \textbf{Max. $M_B$ ($M_\odot$)} \\
\hline
Nuc $S = 0$ & 2.57 & 3.08 \\
Nuc $S = 2$ & 2.59 & 3.07 \\
Nuc $S = 1$, $Y_{\text{lep}} = 0.4$ & 2.54 & 2.86 \\
\\ 
Hyp $S = 0$ & 2.24 & 2.62 \\
Hyp $S = 2$ & 2.22 & 2.53 \\
Hyp $S = 1$, $Y_{\text{lep}} = 0.4$ & 2.34 & 2.68 \\
\hline
\end{tabular}
\label{tab:maximum_masses}
\end{table}

In the upper panel of the Fig. \ref{fig:amr_am_and_amr_bm}, we show the mass-radius distributions with 90\% CI for various compositions. The middle curves in each band represent the median of the mass distributions corresponding to a given radius. Results for two kinds of hadronic matter, without (plain bands) and with (patterned bands) hyperons, are shown. The light red, yellow and pink uniform bands  represent the  90\% CI mass-radius distributions for nucleonic  matter, respectively, at fixed entropy per baryon $S=0$, and $S=2$ for neutrino free matter and $S=1$ for matter with trapped neutrinos i.e. $Y_{\rm lep}=0.4$. The patterned light green, blue, and black regions represent the mass-radius distributions 90\% CI for hyperonic matter for the same entropy per baryon. For both types of matter, the radius of \ac{pns} is larger for higher entropy per baryon, and this is because the larger the entropy, the larger the temperature and the thermal pressure increases. This affects, in particular, the low and medium mass \ac{ns} with a lower central baryonic density. For reference we also included  the astrophysical observations that constrain the $T=0$ \ac{eos} in the upper panel \cite{LIGOScientific:2018cki, Miller:2019cac, Miller:2021qha, Riley:2019yda}. The zero temperature \ac{ns} datasets are compatible with the current astrophysical observations. In the lower panel,  we show the baryonic mass distribution with varying radius of isentropic compact stars ($S=0$, $S=1$ with $Y_{\rm lep}=0.4$, and $S=2$) for both sets of \ac{eos}s. The middle curve in each region represents the median of the \ac{ns} baryonic mass distribution at a given radius. In this figure, colors of both panels have the same meaning. Some comments are in order: i) for matter without hyperons, the maximum gravitational mass or baryonic mass does not change much from the $T=0$ scenario to the different finite $T$ scenarios if  neutrino free matter is considered. However, for matter with trapped neutrinos there is a noticeable decrease on the maximum baryonic mass of the \ac{pns} can stand (see the pink band), although the gravitational mass shows a much smaller decrease with respect to the cold \ac{ns}. This is because nuclear matter with trapped neutrinos is more symmetric, which makes the \ac{eos} softer; 2) matter with hyperons behaves differently: the maximum gravitational and baryonic masses are reached for matter with trapped neutrinos and an entropy per baryon $S=1$. This is due to the fact that a large lepton fraction inhibits the formation of hyperons. However, once the neutrinos escape, the thermodynamic conditions favour the appearance of hyperons, which weaken the \ac{eos} and do not allow such large maximum masses.  This last fact has important consequences: if during the neutrino trapping phase the \ac{pns} has a mass close to its maximum possible mass, then during deleptonisation, given that the baryonic mass is held constant, i.e. no mass is ejected, the star may decay into a low-mass black hole. Some of the \ac{ns} properties for the different scenarios are given in Table \ref{tab:s0_s1_s2}. In particular, we give the median values and 90\% CI limits for the maximum gravitational mass, its radius and the central baryonic energy density and pressure, as well as the radius of \ac{ns} with masses 1.2, 1.4, 1.8, 2.0~$M_\odot$.

\begin{figure*}
    \centering
    \includegraphics[width=0.42\linewidth]{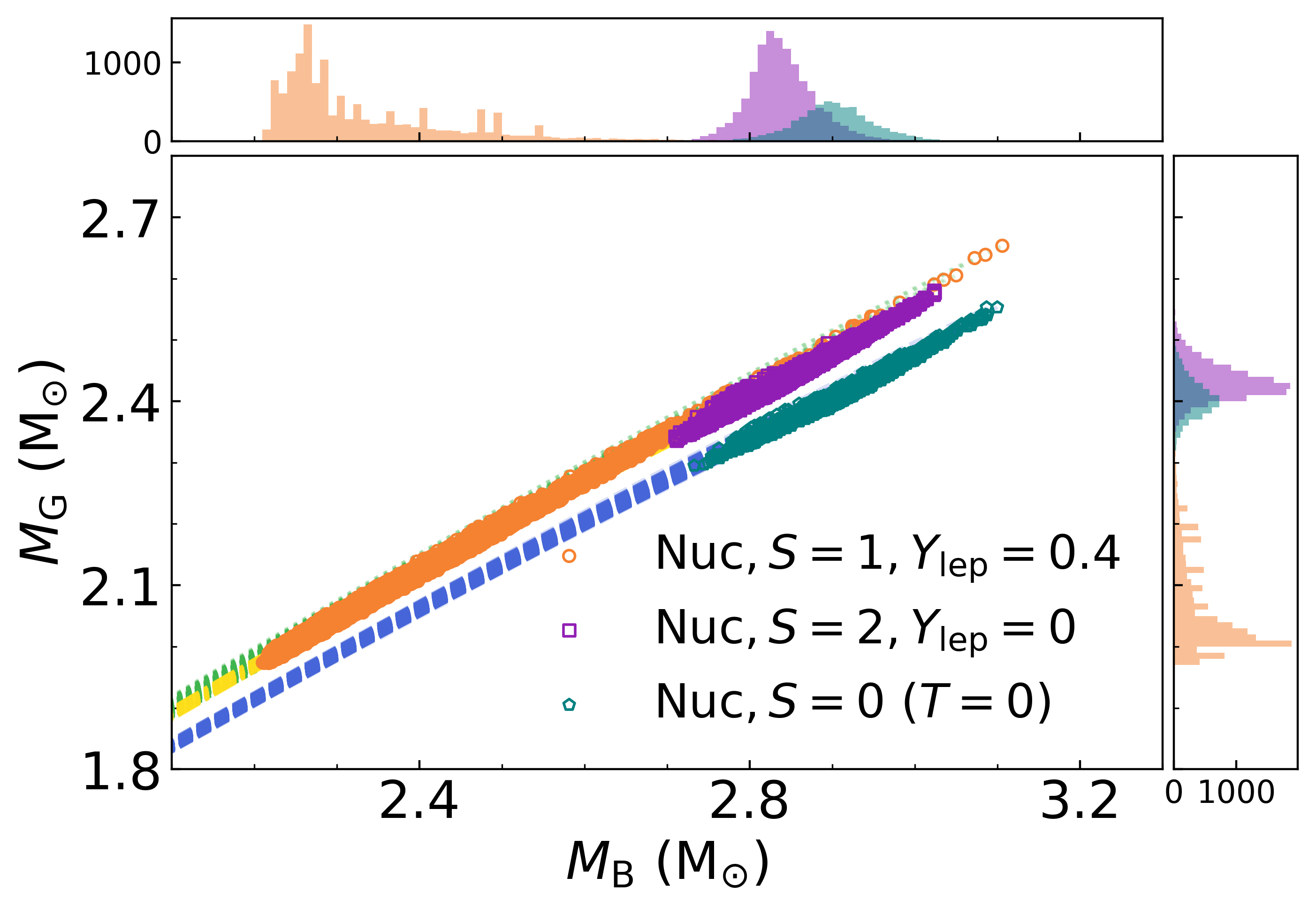}
    \includegraphics[width=0.42\linewidth]{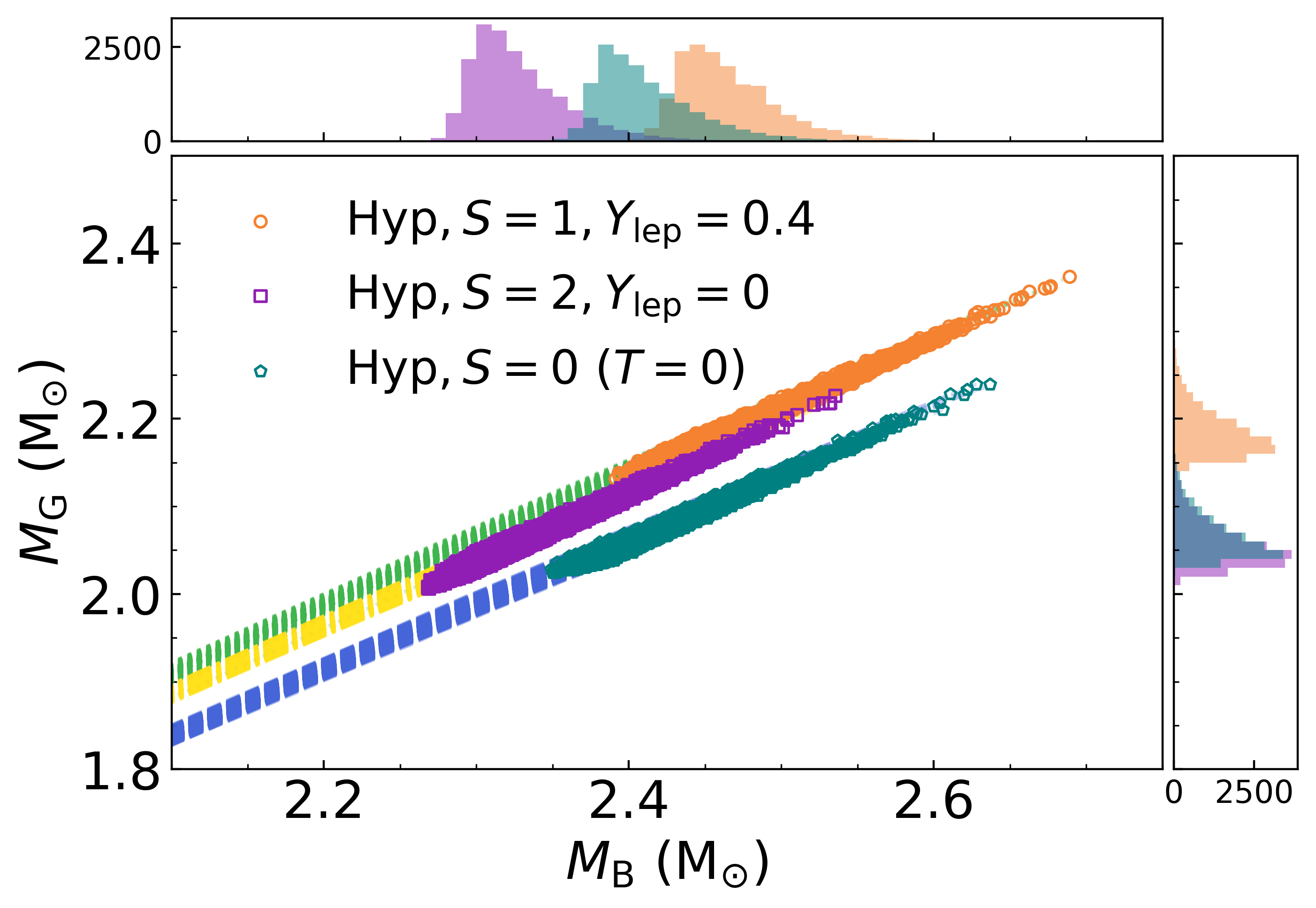}
    \caption{Baryonic versus gravitational masses distribution for (proto) neutron stars for (i) $S = 1,\ Y_{\rm lep} = 0.4$, (ii) $S = 2,\ Y_{\rm lep} = 0$, and (iii) $S=0\ (T=0)$ with (without) hyperonic matter. The open circles denote the maximum mass neutron stars for each \ac{eos}. The distributions of the maximum masses ($M_{\rm G}$ and $M_{\rm B}$) are also displayed on the opposite axes.
    \label{fig:MBvsMG}}
-\end{figure*}

In fact, as discussed in the introduction, \ac{ns}s that contain extra degrees of freedom besides nucleons, e.g. hyperons, $\Delta$'s or a kaon condensate, may become unstable after deleptonization because the extra thermal pressure due to the deposition of the neutrino energy in the star before leaving the \ac{ns} is not enough to stabilize the stars with the largest baryonic masses \cite{Bombaci:1996}. As a result, after the supernova, there will be a delayed collapse into a black hole, possibly with the emission of a $\gamma$-ray burst \cite{Dessart:2007eg, Berezhiani:2002ks, Bombaci:2000cv}. To analyse this possible scenario, we show in Fig. \ref{fig:MBvsMG} the gravitational mass ($M_{\rm G}$) \cite{Kalogera:1996ci} as a function of the baryonic mass ($M_{\rm B}$) \cite{Bombaci:1996} for the three different stages of the \ac{pns} evolution \cite{Rabhi:2011ej} and for all the \ac{eos}s of the both data sets: the nucleonic and the hyperonic data sets. As three different snapshots of the \ac{pns} evolution, we consider here: i) a first stage with trapped neutrinos, a lepton fraction $Y_{\rm lep} = 0.4$ and entropy per particle $S=1$ ($S=1,\ Y_{\rm lep} = 0.4$); ii) a second stage when neutrinos leave the star after depositing their energies, the temperature of the star increases so that the entropy per particle becomes $2$ ($S=2,\ Y_{\nu_e} = 0$); iii) a third and final stage when the star becomes cold and stable \ac{ns} $S=0\ (T=0),\ Y_{\nu_e} = 0$.

\begin{figure*}
    \centering
    \includegraphics[width=0.42\linewidth]{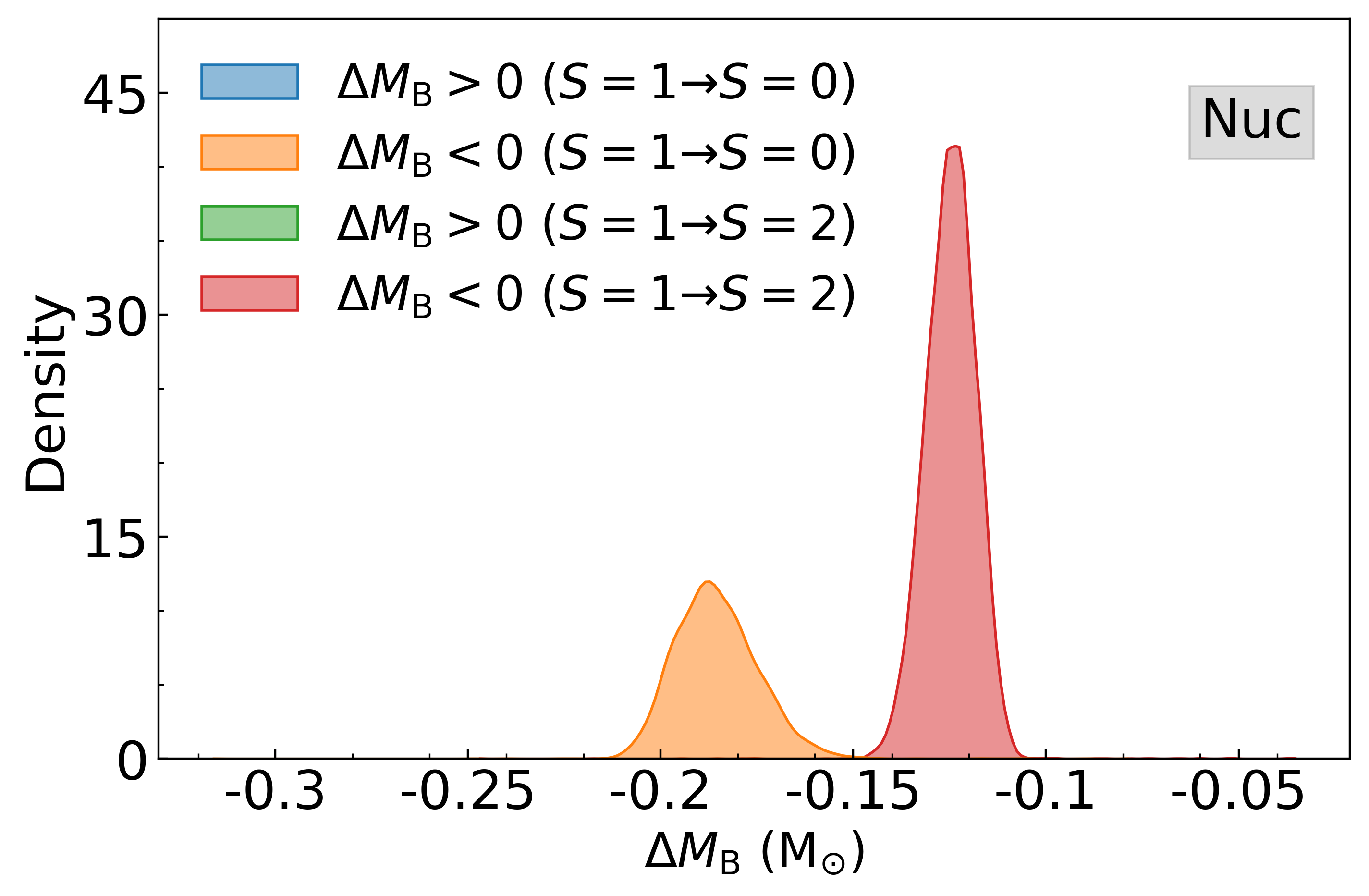}
    \includegraphics[width=0.42\linewidth]{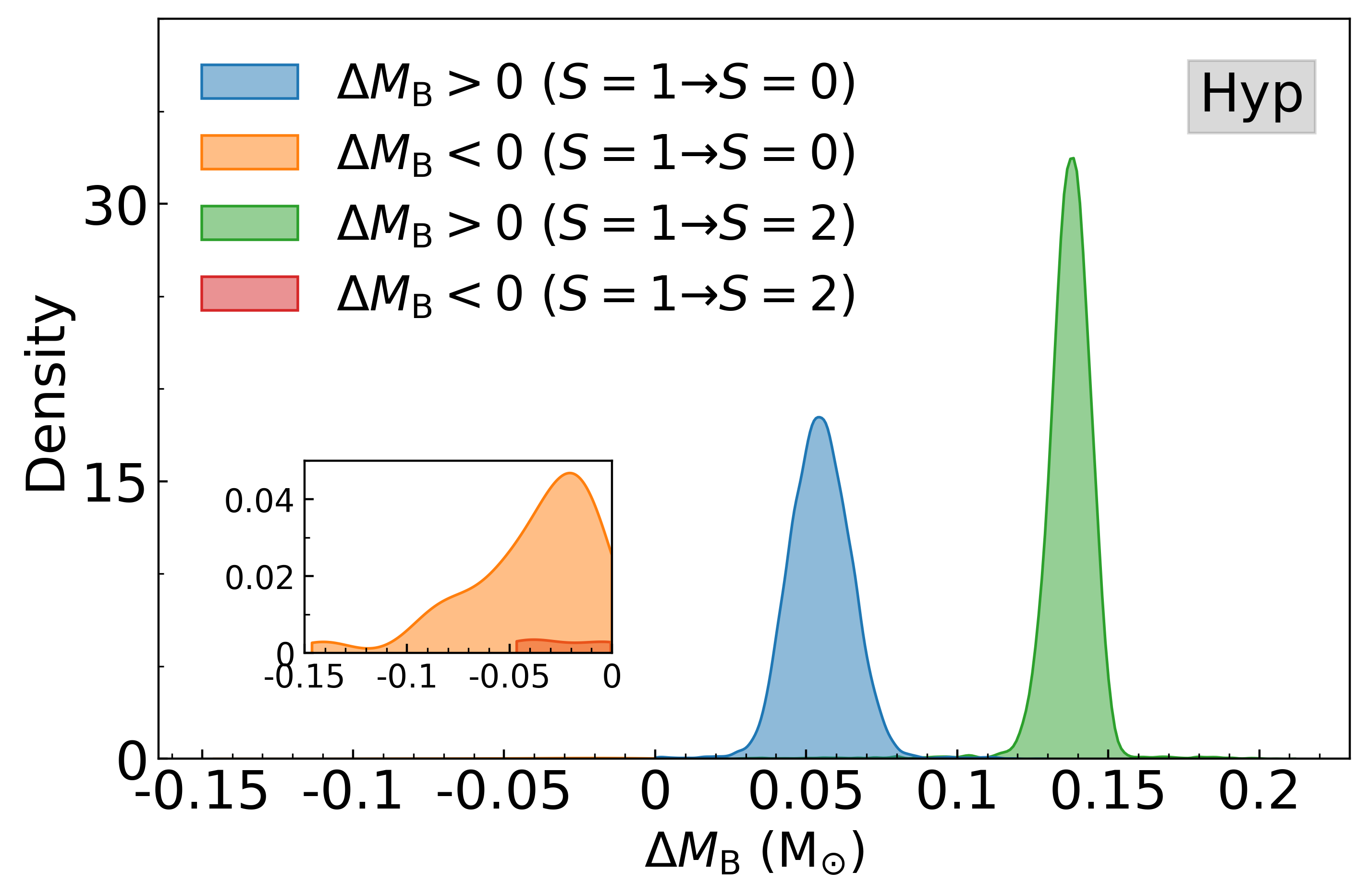}
    \caption{The frequency plots along with the distribution functions of $\Delta M_{\rm B} = M_{\rm B}(S=1,Y_{\rm lep}=0.4) - M_{\rm B}(S=0,2)$ for both cases, like (i) $S=1,\ Y_{\rm lep}=0.4\ \to\ S=0$, and (ii) $S=1,Y_{\rm lep}=0.4\ \to\ S=2$ in the cases of nuclear matter and hyperonic matter. In the case of nuclear matter, the stars are spread over a large (x) range while in the case hyperonic matter, they are concentrated in a small range.}
    \label{fig:dist_dmb_npe_and_hyp}
\end{figure*}

In Fig. \ref{fig:MBvsMG}, the different lines from the top to bottom correspond to stages (i), (ii) and (iii), respectively. The different markers (orange circle, magenta square and pentagon teal) represent the localization of the maximum mass configurations. We also show on the top and right of each panel the distribution of the maximum mass configurations on the baryonic mass ($M_{\rm B}$) and gravitational mass ($M_{\rm G}$). In Table \ref{tab:maximum_masses}, we summarize the maximum baryonic and gravitational mass in each scenario. With the nucleonic dataset, the maximum baryonic mass at stage $S=1,Y_{\rm lep}=0.4$ \ac{pns} spreads over a range of values much smaller than the maximum baryonic mass of a cold star or a deleptonized warm \ac{pns}. Although with a narrower distribution, the neutrinoless mass distribution at stage $S=2$ takes smaller values than the corresponding distribution of the cold stars. These results indicate that the star will evolve to a stable cold star. This can be confirmed looking at the distribution functions plotted in the left panel of Fig. \ref{fig:dist_dmb_npe_and_hyp} where the differences $\Delta M_{\rm B}(1\to0) = M_{\rm B}(S=1,Y_{\rm lep}=0.4) - M_{\rm B}(S=0) $ and $\Delta M_{\rm B}(1\to2)= M_{\rm B}(S=1,Y_{\rm lep}=0.4) - M_{\rm B}(S=2) $ are shown. From the first stage ($S=1,\ Y_{\rm lep} = 0.4$) represented by the green distribution to the last one ($S=0$, blue distribution), the maximum baryonic mass may increase as $\sim 0.08 M_\odot$. In Table \ref{tab:fwhm}, we give the full width at half maxima for the distributions, to complement the information in the figure. 

\begin{table}[h]
\centering
\caption{Full width at half maxima for the distributions showing in the Fig. \ref{fig:dist_dmb_npe_and_hyp}.}
\begin{tabular}{|r|cc|cc|}
\toprule
& \multicolumn{4}{c|}{FWHM} \\ \cline{2-5}
& \multicolumn{2}{c|}{$S=1,Y_{\rm lep} = 0.4 \to S=0$} & \multicolumn{2}{c|}{$S=1,Y_{\rm lep}=0.4 \to S=2$} \\ \cline{2-5}
& \multicolumn{1}{c}{$\Delta M_{\rm B} > 0$} & \multicolumn{1}{c|}{$\Delta M_{\rm B} < 0$} & \multicolumn{1}{c}{$\Delta M_{\rm B} > 0$} & \multicolumn{1}{c|}{$\Delta M_{\rm B} < 0$} \\ \cline{1-5}
\multicolumn{1}{|c|}{Nuc} & \multicolumn{1}{c}{0} & \multicolumn{1}{c|}{0.025} & \multicolumn{1}{c}{0} & \multicolumn{1}{c|}{0.017} \\
\multicolumn{1}{|c|}{Hyp} & \multicolumn{1}{c}{0.03} & \multicolumn{1}{c|}{0.051} & \multicolumn{1}{c}{0.015} & \multicolumn{1}{c|}{0.076}\\
\toprule
\end{tabular}%
\label{tab:fwhm}%
\end{table}%

A different evolution occurs for stars that include non-nucleonic degrees of freedom as shown in the right panels in Figs. \ref{fig:MBvsMG} and \ref{fig:dist_dmb_npe_and_hyp}. As may be observed from the right panel of Fig. \ref{fig:MBvsMG}, for the first stage ($S=1,\ Y_{\rm lep}=0.4$), the maximum gravitational mass ($M_{\rm G}$) ranges from 2.15 $M_{\odot}$ to 2.34 $M_{\odot}$. On the other hand, for the second and the third stages the maximum gravitational mass upper limits are smaller compared to the same for stage (i). Further, it may be noted that the transition from the first stage to the cold star is accompanied by an average reduction of the mass of $\sim 0.1\ M_{\odot}$. In fact, there is also a noticeable reduction of the maximum baryonic mass with deleptonization, i.e. in the transition of stage (i) to stage (ii), see Table \ref{tab:maximum_masses}. Of course, only a simulation of the evolution may give more concrete conclusions, but our results seem to indicate that if hyperons enter the composition of \ac{ns}s, the destabilization of the \ac{pns} occurs in an early stage of the \ac{pns} evolution for stars at stage (i) ($S=1,\ Y_{\rm lep}=0.4$) with gravitational mass beyond 2.2 $M_{\odot}$ becoming unstable as no stable solution exists with the same baryonic mass for star at stage (ii) ($S=2, Y_{\nu_e} = 0$). This could mean that compact objects with a gravitational mass above $\sim 2.2\, M_\odot$, will collapse to a black hole accompanied with $\gamma$-ray burst \cite{Dessart:2007eg}.

\section{Summary and Conclusions} \label{sec:summary_and_conclusion}
In this study, we have investigated the evolution of \ac{pns}s that incorporate both nucleonic and hyperonic degrees of freedom within a relativistic mean-field framework considering two constrained datasets: one for nucleonic matter and the second for hyperonic matter, each set with 18000 \ac{eos}. Both the datasets were obtained within a Bayesian inference calculation that imposed nuclear matter and observational constraints. These data sets satisfy the two solar mass constraint for the lower limit of the maximum \ac{ns} mass, and, in addition, constraints from neutron matter chiral effective field theory calculations and several nuclear matter properties at saturation density (saturation density, binding energy, incompressibility and symmetry energy) at zero temperature. Our analysis, reveals several key findings with significant astrophysical implications:
\begin{itemize}
    \item First, the presence of hyperons fundamentally alters the thermal and structural evolution of \ac{pns}s. When hyperons are included in the stellar composition, they significantly soften the \ac{eos}, leading to a substantial reduction in the maximum supportable baryonic mass during deleptonization and cooling. This confirms the results of \cite{Brown:1993jz, Glendenning:1994za, Prakash:1996xs} where it is shown that hyperonic \ac{pns}s with a mass close to the maximum mass configurations in the \ac{pns} stage with trapped neutrinos are susceptible to delayed collapse into black holes, while this is not expected to occur in  purely nucleonic \ac{ns}, except in case of  post bounce accretion.

    \item Second, our thermal analysis demonstrates that the thermal adiabatic index ($\Gamma_{\rm Th}$) exhibits distinctive behavior in hyperonic matter, showing significant fluctuations and a characteristic dip coinciding with the onset of hyperons, as discussed in \cite{Kochankovski:2022rid}. This reflects the redistribution of thermal energy across additional degrees of freedom, resulting in reduced thermal pressure support. The nucleonic \ac{eos}, on the contrary, maintains a smooth thermal profile throughout the evolution. The band that defines ($\Gamma_{\rm Th}$) was shown to be quite narrow at a 90\% CI above $T=10$ MeV - of the order of 0.1 - reflecting the fact that the role of the interaction becomes negligible and the behavior of the \ac{eos} is defined by the thermal effects and the number of degrees of freedom.  For hyperonic matter, the corresponding $\Gamma_{\rm Th}$ band is also quite narrow above $\sim 30$ MeV. However, at $T=10$ MeV this band has a width $\gtrsim 1.5$ above the hyperon onset density at zero temperature and up to the central density. This may have a non-negligible effect on the \ac{ns}
    merger evolution.
\end{itemize}

Our gravitational-baryonic mass analysis offers particularly compelling evidence for delayed collapse scenarios. \ac{pns}s with trapped neutrinos ($S=1$, $Y_{\rm lep} = 0.4$) including hyperons can temporarily support larger baryonic masses than their cold, deleptonized counterparts. This metastability occurs in hyperonic models, where the transition from lepton-rich to neutrino-free states results in an average reduction of $\sim 0.1$ $M_\odot$ in the maximum supportable gravitational and baryonic masses. Our analysis suggests that if hyperons are indeed present in \ac{ns}s, compact objects with gravitational masses exceeding 2.2 $M_\odot$ may actually be black holes formed through delayed collapse, and the maximum gravitational mass of a \ac{ns} with hyperons in its core can not exceed $\sim 2.2 M_\odot$. The last limit rises to $\sim 2.5\ M_\odot$ for the nucleonic star.

These results have profound implications for multi-messenger astronomy. The delayed collapse mechanism could explain certain features in supernova neutrino signals, possibly related to the production of gamma-ray bursts in failed supernovae or the cessation of the neutrino signal when a black hole forms as suggested in \cite{Prakash:1996xs}. In addition, our results suggest that current maximum mass measurements of \ac{ns}s may indirectly constrain the occurrence of hyperonic matter in their cores, with maximum masses above 2.2 $M_\odot$ indicating the absence of hyperons. It would be interesting to investigate how these results might be strengthened if the \ac{pns} contains dark matter (DM). The presence of DM leads to larger baryon densities in the \ac{ns} or \ac{pns} centers, allowing larger hyperon abundances for \ac{ns} of the same mass, and thus leading to stronger deleptonization effects.  This could imply that in environments with high DM content, the decay occurs during deleptonization of \ac{pns} into black holes for \ac{pns} with masses below $\sim$2.2~M$_\odot$.

Future work in this direction should focus on time-dependent simulations of this evolution process, incorporating detailed neutrino transport mechanisms and more sophisticated treatment of phase transitions. Continued observations of massive neutron stars, supernova neutrino signals, and gravitational waves from merger events will provide critical tests of these theoretical predictions, potentially shedding light on the fundamental composition of matter under extreme conditions.\\

\begin{acknowledgments}
D.K. acknowledges the financial support from the Science and Engineering Research Board (SERB), Govt. of India, under the project grant ``Core Research Grant (CRG/2022/000663)". The authors express their gratitude to the Deucalion HPC in Portugal for its support in the Advanced Computing Project 2025.00067.CPCA A3, RNCA (Rede Nacional de Computação Avançada), financed by the FCT (Fundação para a Ciência e a Tecnologia, IP, Portugal). TM and CP acknowledge the partial supported by funds from FCT (Fundação para a Ciência e a Tecnologia, I.P, Portugal) under projects UIDB/04564/2020 and UIDP/04564/2020, with DOI identifiers 10.54499/UIDB/04564/2020 and 10.54499/UIDP/04564/2020, respectively, and the project 2022.06460.PTDC with the associated DOI identifier 10.54499/2022.06460.PTDC. 
\end{acknowledgments}

\bibliography{referencias}

\end{document}